\documentclass[12pt, a4paper]{article} 

\baselineskip=\normalbaselineskip

\usepackage{mathrsfs,amsbsy,amssymb,latexsym,amsfonts,amsmath,amsthm}
\usepackage{graphicx,color}

\usepackage{bm}

\makeatletter
\catcode`\@=11
\@addtoreset{equation}{section}

\def\@seccntformat#1{\csname the#1\endcsname.~~}
\makeatother

\begin{document}
\begin{titlepage}
\renewcommand{\thefootnote}{\fnsymbol{footnote}}
\begin{flushright}
KUNS-2732
\end{flushright}
\vspace*{1.0cm}

\begin{center}
{\Large \bf 
  Emergence of AdS geometry \\
  in the simulated tempering algorithm
}
\vspace{1.0cm}

\centerline{
{Masafumi Fukuma${}^1$}%
\footnote{E-mail address: 
fukuma@gauge.scphys.kyoto-u.ac.jp},  
{Nobuyuki Matsumoto${}^1$}%
\footnote{E-mail address: 
nobu.m@gauge.scphys.kyoto-u.ac.jp} and
{Naoya Umeda${}^2$}%
\footnote{E-mail address: 
n\_umeda@gauge.scphys.kyoto-u.ac.jp}%
}


\vskip 0.8cm
${}^1${\it Department of Physics, Kyoto University, Kyoto 606-8502, Japan}\\
\vskip 0.1cm
${}^2${\it PricewaterhouseCoopers Aarata LLC, \\
Otemachi Park Building, 1-1-1 Otemachi, Chiyoda-ku, Tokyo 100-0004, Japan}
\vskip 1.2cm 

\end{center}

\begin{abstract}
In our previous work \cite{FMU}, 
we introduced to an arbitrary Markov chain Monte Carlo algorithm 
a distance between configurations. 
This measures the difficulty of transition 
from one configuration to the other, 
and enables us to investigate the relaxation of probability distribution 
from a geometrical point of view. 
In this paper, we investigate the global geometry of a stochastic system 
whose equilibrium distribution is highly multimodal 
with a large number of degenerate vacua. 
We show that, 
when the simulated tempering algorithm is implemented to such a system, 
the extended configuration space has 
an asymptotically Euclidean anti-de Sitter (AdS) geometry. 
We further show that this knowledge of geometry 
enables us to optimize the tempering parameter 
in a simple, geometrical way.

\end{abstract}
\end{titlepage}

\pagestyle{empty}
\pagestyle{plain}

\tableofcontents
\setcounter{footnote}{0}

\section{Introduction}
\label{sec:introduction}

Let $\mathcal{M}=\{x\}$ be a configuration space, 
and $S(x)$ an action. 
We are often concerned with calculating 
the vacuum expectation value (VEV) of an operator $\mathcal{O}(x)$ 
with respect to the action:
\begin{align}
 \langle \mathcal{O}(x) \rangle
 \equiv \frac{1}{Z}\,\int dx\,e^{-S(x)}\,\mathcal{O}(x)
 \quad
 \Bigl(Z=\int dx\,e^{-S(x)}\Bigr). 
\end{align}
In a Markov chain Monte Carlo (MCMC) simulation, 
we set up an algorithm that generates a stochastic process 
$p_n(x) \to p_{n+1}(x)=\int dy\,P(x|y)\,p_n(y)$ 
such that the probability distribution $p_n(x)$ relaxes 
to the desired equilibrium distribution 
$e^{-S(x)}/Z$ in the limit $n\to\infty$. 
In order for such an algorithm to be practically useful, 
the relaxation to equilibrium needs to be sufficiently rapid. 
However, when the equilibrium distribution is multimodal 
(i.e., when the action has very high potential barriers), 
transitions between configurations belonging to different modes 
take extraordinarily long computational times, 
which delay the relaxation to equilibrium 
and make the MCMC simulation almost impractical.

To accelerate the transitions, 
there have been invented various methods, 
including the overrelaxation \cite{Creutz:1987xi} 
and the simulated and the parallel tempering methods 
\cite{Marinari:1992qd,Swendsen1986,Geyer1991,Earl2005}. 
In the simulated tempering method \cite{Marinari:1992qd}, for example, 
as will be briefly reviewed in subsection \ref{sec:sim_temp_alg}, 
one picks up a parameter $\beta$ of the model 
(such as the overall coefficient of the action) 
as a \textit{tempering parameter}, 
and extends the configuration space 
by treating $\beta$ as an additional dynamical variable. 
We then design a MCMC algorithm 
such that configurations belonging to different modes 
for the original action 
now can be connected easily 
by passing through configurations 
in the extended configuration space. 
In order for such an algorithm to work, 
transitions along the $\beta$ direction must have significant acceptance rates, 
which enforces us to make a careful adjustment of the parameters 
that are additionally introduced when extending the configuration space.

In our previous work \cite{FMU}, 
we introduced to an arbitrary MCMC algorithm 
the notion of {\em distance between configurations}. 
This measures the difficulty of transition 
from one configuration to the other, 
and enables us to investigate the relaxation of probability distribution 
from a geometrical point of view. 
We also introduced the {\em coarse-grained configuration space} 
to a highly multimodal stochastic system, 
where configurations in the same mode are regarded as a single configuration 
\cite{FMU}. 
This perspective facilitates the understanding of 
the global geometry of the configuration space, 
and is justified by the fact that 
distances between different modes take so large values
that the difference of distances between two configurations 
in the same mode can be effectively neglected.  
In this paper, 
we investigate the global geometry of such a highly multimodal stochastic system 
with a large number of degenerate vacua. 
We show that, 
when the simulated tempering algorithm is implemented, 
an asymptotically Euclidean anti-de Sitter (AdS) geometry emerges 
in the extended, coarse-grained configuration space. 
We further show that 
this knowledge of geometry enables us 
to optimize the additional parameters 
in a simple, geometrical way.%
\footnote{
 In our previous paper \cite{FMU} and 
 also in the first draft of the present paper, 
 it was incorrectly claimed that 
 an AdS geometry emerges 
 as a result of the optimization of the additional parameters. 
 Actually, as will be seen in section~\ref{sec:emergence2}, 
 the AdS geometry appears independently of the choice the parameters. 
 However, as will be shown in section~\ref{sec:optim}, 
 this fact in turn can be used to determine 
 the optimized form of the parameters in an easy way. 
} 

This paper is organized as follows. 
In section \ref{sec:model}, 
we give a brief review on the distance between configurations 
for a MCMC algorithm \cite{FMU}. 
We mainly consider a system 
whose equilibrium distribution is highly multimodal 
with a large number of degenerate vacua. 
We also introduce the concept of the coarse-grained configuration space 
and define the metric on the space. 
In section~\ref{sec:emergence2}, 
we implement the simulated tempering method 
by taking the tempering parameter 
to be the overall coefficient of the action, 
and show that 
the geometry of the extended, coarse-grained configuration space 
is given by an asymptotically Euclidean AdS space metric. 
In Section \ref{sec:optim}, 
we optimize the tempering parameter such that the distances get minimized, 
and show that this can be easily done in a geometrical way. 
The conclusion is confirmed by direct numerical calculations. 
Section \ref{sec:conclusion} is devoted to conclusion 
and outlook for future work.

\section{Distance between configurations in MCMC simulations}
\label{sec:model}

In this section, 
we give a brief review on the distance between configurations 
for a MCMC algorithm \cite{FMU}. 
We mainly consider a system 
whose equilibrium distribution is highly multimodal 
with a large number of degenerate vacua. 
We then introduce the coarse-grained configuration space, 
where configurations in the same mode are regarded 
as a single configuration.

\subsection{Definition of the distance between configurations}
\label{sec:def_distance}

Let us consider a configuration space $\mathcal{M} = \{x\}$ 
with a real-valued action $S(x)$. 
Suppose that we make a numerical simulation 
to estimate the VEVs of observables by using a given MCMC algorithm. 
We denote by $P(x|y)=\langle x | \hat{P}\,|y \rangle$ 
the transition probability 
from a configuration $y$ to a configuration $x$ at a single Markov step. 
The Markov property then says that 
the transition probability from $y$ to $x$ at $n$ steps 
is given by $P_n(x|y) = \langle x | \hat{P}^n |y \rangle$. 
We assume that the stochastic process 
has the unique equilibrium distribution $e^{-S(x)}/Z$\ $(Z=\int dx e^{-S(x)})$ 
and $P(x|y)$ satisfies the detailed balance condition 
$P(x|y)\,e^{-S(y)} = P(y|x)\,e^{-S(x)}$. 
We further assume that all the eigenvalues of $P(x|y)$ are positive 
(see \cite{FMU} for mathematical details).

The distance $\theta_n(x,y)$ between two configurations $x,y$ \cite{FMU}
is then defined by 
\begin{align}
  \theta_n(x,y) \equiv \arccos \biggl(
  \frac{P_n(x|y)\, P_n(y|x)}{P_n(x|x)\, P_n(y|y)} \biggr).
  \label{distance_theta}
\end{align}
One can easily show  
that $\theta_n(x,y)$ satisfies the axioms of distance \cite{FMU}: 
\begin{align}
  \bullet& \ \theta_n(x,y)\geq 0, 
\label{distance1}
\\
  \bullet& \ x=y \Leftrightarrow \theta_n(x,y) = 0, 
\label{distance2}
\\
  \bullet& \ \theta_n(x,y) = \theta_n(y,x), 
\label{distance3}
\\
  \bullet& \ \theta_n(x,y) + \theta_n(y,z) \geq \theta_n(x,z),
\label{distance4}
\end{align}
and vanishes in the limit $n\to\infty$ 
for any two configurations $x,\,y\in\mathcal{M}$: 
\begin{align}
 \lim_{n\to\infty}\,\theta_n(x,y) = 0,
\end{align}
because $P_n(x|y)\to e^{-S(x)}/Z$ in the limit $n\to\infty$. 
The distance $\theta_n(x,y)$ actually measures 
the difficulty of transition from $y$ to $x$ at $n$ steps 
in the sense that: 
\begin{itemize}
\item[\textbullet]
  When configurations $x$ and $y$ belong to different modes 
  of the equilibrium distribution, 
  $\theta_n(x,y)$ is large for a finite $n$. 
\item[\textbullet]
  If $x$ can be easily reached from $y$, 
  $\theta_n(x,y)$ is small even for a finite $n$. 
\end{itemize}
One can further show that, 
as long as the chosen algorithm generates only local moves of configuration,
the large scale structure of distance $\theta_n(x,y)$ takes a universal form, 
in the sense that differences of distance between two such algorithms 
can always be absorbed into a rescaling of $n$ \cite{FMU}.

In \cite{FMU} 
we introduced a few distances in addition to $\theta_n(x,y)$, 
among which is the distance $d_n(x,y)$ 
that is defined by 
\begin{align}
  d_n(x,y) = \sqrt{-\log\biggl(
  \frac{ P_n(x|y)\, P_n(y|x)}{P_n(x|x)\, P_n(y|y)} \biggr)}
\label{distance_d_P}
\end{align}
and is related with $\theta_n(x,y)$ as  
\begin{align}
  \cos \theta_n(x,y) = e^{-(1/2)\,d_n^2(x,y)}.
\label{distance_d}
\end{align}
$d_n(x,y)$ agrees with $\theta_n(x,y)$ when they take small values,  
and satisfies almost the same properties as $\theta_n(x,y)$. 
The only exception is that $d_n(x,y)$ generically does not satisfy 
the triangle inequality for the original configuration space 
$\mathcal{M}$. 
However, as will be discussed in subsection \ref{sec:prop}, 
$d_n(x,y)$ is more useful than $\theta_n(x,y)$ 
for investigating the large scale geometry of the configuration space 
and does satisfy the triangle inequality 
when the configuration space is coarse-grained. 
Thus, we will mainly use $d_n(x,y)$ in the following discussions.

\subsection{Coarse-grained configuration space}
\label{sec:coarse_grained_def}

For a stochastic system 
whose equilibrium distribution 
is highly multimodal with a large number of degenerate vacua, 
distances $d_n$ between two different modes take so large values 
that the difference of distances between two configurations 
in the same mode can be effectively neglected. 
This leads us to introduce the coarse-grained configuration space 
$\bar{\mathcal{M}}$
by regarding configurations in the same mode as a single configuration 
\cite{FMU}. 
In the following, we assume that a way of separation between different modes 
is uniform and translationally invariant in $\bar{\mathcal{M}}$. 
A typical model which has this property 
can be given by the action%
\footnote{
 The universality of our distance 
 (see \cite{FMU} and subsection \ref{sec:def_distance} of the present paper) 
 indicates that 
 the large scale geometry of any multimodal system 
 with a large number of degenerate vacua 
 can be expressed by the action \eqref{cosine_action}
 if the local minima are distributed in a uniform way. 
} 
\begin{align}
  S(\mathbf{x};\beta_0)=\beta_0\,\sum_{\mu=1}^D \,[1-\cos(2\pi x_\mu) ].
\label{cosine_action}
\end{align}
Here, the configuration space is a $D$-dimensional torus 
with the period $2L$ for every direction: 
$\mathcal{M}=
\{\mathbf{x}= (x_\mu)\,|\,{-L} < x_\mu \leq L \ (\mu = 1,\ldots,D) \}$ 
(we impose the periodic boundary condition). 
This action certainly gives a multimodal equilibrium distribution 
when $\beta_0\gg1$. 
The coarse-grained configuration space $\bar{\mathcal{M}}$ 
is given by the $D$-dimensional lattice torus 
$\bar{\mathcal{M}} = \{n=(n_\mu) \,|\, n_\mu={-L}+1,\ldots,L-1,L \}$, 
that consists of the degenerate classical vacua  
and has a translational invariance.

\subsection{Distance on the coarse-grained configuration space}
\label{sec:prop}

We here explain why we think that $d_n(x,y)$ is more suitable than $\theta_n(x,y)$ 
for investigating the geometry of $\bar{\mathcal{M}}$.

Firstly, $d_n(x,y)$ has a better resolution 
on the large scale structure of configuration space. 
In fact, as can be seen from its definition, 
when transitions between two configurations $x$ and $y$ 
happen only very rarely, 
$d_n(x,y)$ can take a large value without limitation, 
while the value of $\theta_n(x,y)$ is saturated close to $\pi/2$.

Secondly, the distance $d_n(x,y)$ gives a flat Euclidean metric 
for a simple Gaussian distribution \cite{FMU}. 
In fact, for a quadratic action 
$S(\mathbf{x}) = (\omega/2)\,\mathbf{x}^2$ 
defined on the configuration space 
$\mathcal{M}=\{\mathbf{x}\,|\,\mathbf{x}\in\mathbb{R}^D\}$ 
with the Langevin algorithm, 
the distance between $\mathbf{x}$ and $\mathbf{y}$ is given by 
\begin{align}
  d_n(\mathbf{x},\mathbf{y})
  = \sqrt{\frac{\omega}{2\sinh(\omega n\epsilon)}}\,|\mathbf{x}-\mathbf{y}|,
\label{distance_Gaussian}
\end{align}
where $\epsilon$ is the step size of the discretized fictitious time 
in the Langevin equation. 
$\theta_n(x,y)$, to the contrary, has a complicated expression.

\begin{figure}[t]
  \centering
  \includegraphics[width=7.5cm]{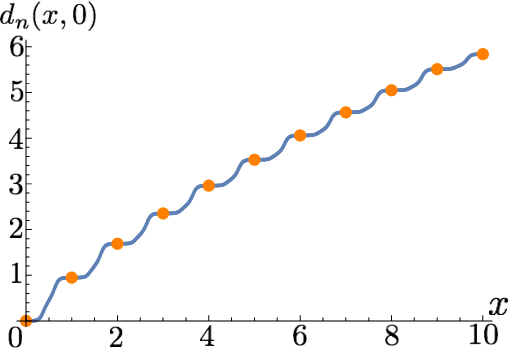}
  \caption{$d_n(x,0)$ for $0\leq x \leq L$. 
  We set $n=2,500$ , $\beta_0=3$ and $L=10$.}
  \label{fig:triangled1}
\end{figure}

Lastly, the distance $d_n(x,y)$ 
is expected to satisfy the triangle inequality 
on the coarse-grained configuration space $\bar{\mathcal{M}}$. 
Although we still do not have a rigorous proof on this statement, 
we can check it numerically for the model \eqref{cosine_action}. 
To see this, 
we first note that the triangle inequality in $\bar{\mathcal{M}}$ 
is equivalent to the concavity of $d_n(x,y)$ as a function of $r=|x-y|$ 
($x,y\in\bar{\mathcal{M}}$) 
due to the translational symmetry of this model. 
Figure \ref{fig:triangled1} shows 
$d_n(x,0)$ for the action \eqref{cosine_action} 
with $D=1$, $L=10$ and $\beta_0=3$. 
%
As a MCMC algorithm, 
we use the Metropolis algorithm 
with the Gaussian proposal distribution with standard deviation 
$\sigma_x = 1.6/(2 \pi \sqrt{\beta_0})$.%
\footnote{
 See footnote~\ref{fn:sigmax} about how we choose the value of $\sigma_x$.
} 
We discretize the original configuration space $\mathcal{M}$ 
with spacing $0.02$, 
and numerically construct the transition matrix $P(x|y)$. 
We then calculate $d_n(x,0)$ for $0 \leq x \leq L$ with $n=2,500$. 
We see that, although $d_n(x,0)$ does not satisfy concavity 
for generic configurations $x\in\mathcal{M}$, 
it becomes a concave function of $r=|x|$ 
when $x$ is on the lattice of the coarse-grained configuration space 
(drawn as orange points in the figure).

We now define the metric $ds^2$ 
in the coarse-grained configuration space $\bar{\mathcal{M}}$ as
\begin{align}
 ds^2 
 \equiv d_n^2(x,\,x+dx),
\label{distance}
\end{align}
where $x$ and $x+dx$ are nearby points residing on $\bar{\mathcal{M}}$
\cite{FMU}. 
For the translationally invariant, parity even action \eqref{cosine_action}, 
the metric will take the following form for $\beta_0\gg1$:
\begin{align}
  ds^2 = {\rm const.}\, \beta_0\, \sum_{\mu=1}^D\,dx_\mu^2,
\label{coarse-grained_metric}
\end{align}
because the transition probability between two neighboring modes 
is given by $O(e^{-{\rm const.}\,\beta_0})$ 
as can be easily seen by an instanton calculus \cite{FMU}. 

\section{Emergence of the Euclidean AdS geometry}
\label{sec:emergence2}

In this section, 
we implement the simulated tempering method for the action \eqref{cosine_action} 
by taking the tempering parameter 
to be the overall coefficient of the action, $\beta_0$. 
The configuration space is thus extended 
so that the tempering parameter is also treated as a dynamical variable. 
We show that 
the global geometry of the extended configuration space 
is given by an asymptotically Euclidean AdS metric. 
In subsection~\ref{sec:sim_temp_alg} 
we briefly explain the simulated tempering algorithm, 
and in subsection~\ref{sec:u_approx} 
we write down the transition matrix explicitly for large $\beta$. 
In subsection~\ref{sec:metric} 
we prove that the geometry of the extended, coarse-grained configuration space 
is given by an AdS metric for large $\beta$. 
This statement is numerically verified in subsection~\ref{sec:numerical}.

\subsection{Simulated tempering algorithm}
\label{sec:sim_temp_alg}

We here give a brief review of the simulated tempering algorithm 
\cite{Marinari:1992qd}. 
As is commented in Introduction, 
this is introduced to a multimodal stochastic system 
in order to accelerate the relaxation to global equilibrium.

A MCMC simulation proceeds as follows: 
\begin{enumerate}
\item
We choose a \textit{tempering parameter} (to be denoted by $\beta_0$) 
from parameters in the action, 
which we write as $S(x;\beta_0)$ 
to indicate its dependence on $\beta_0$. 

\item
We introduce the tempering parameter set 
$\mathcal{A} = \{\beta_a\}$ $(a=0,1,\ldots,A)$, 
to which belongs the original parameter $\beta_0$. 

\item
We extend the original configuration space $\mathcal{M}=\{x\}$ 
to $\mathcal{M}\times\mathcal{A}=\{X=(x,\beta_a)\}$. 

\item
We set up a Markov chain on the extended configuration space 
such that the global equilibrium distribution $P_{\rm eq}(X)$ 
takes the form%
\footnote{ 
  The weight $w_a$ will be chosen as 
  \begin{align}
   P_\mathrm{eq}(X)=\frac{1}{(A+1)Z(\beta_a)}\,e^{-S(x;\beta_a)} 
   \quad \Bigl(Z(\beta_a) = \int dx \,e^{-S(x;\beta_a)}\Bigr)
  \nonumber
  \end{align}
  in the following discussion 
  to make configurations with $\beta_a$ appear 
  with the same appearance ratio for all $a$. 
} 
\begin{align}
 P_{\rm eq}(X) = P_{\rm eq}(x,\beta_a) = w_a\, e^{-S(x;\beta_a)}.
\label{eq_sim_temp}
\end{align}

\item
After the system is well regarded as reaching global equilibrium, 
we retrieve a subsample with $\beta_{a=0}$ 
out of a full sample taken from the extended configuration space, 
and estimate the VEVs by sample averages with respect to the subsample. 
\end{enumerate}
In this paper, 
we realize the global equilibrium~(\ref{eq_sim_temp})  
by repeating the following steps: 
\begin{itemize}
\item[(1)] We generate a local move in the $x$ direction, 
$X = (x,\beta_a)\to X' = (x',\beta_a)$, 
with the Metropolis algorithm. 
This is repeated $2k$ times so that local equilibrium is realized 
for fixed $\beta_a$. 

\item[(2)] We generate a local move in the $\beta$ direction, 
$X = (x,\beta_a)\to X' = (x,\beta_{a\pm 1})$, 
using the Metropolis algorithm, 
where an adjacent tempering parameter is proposed with probability $1/2$ 
and accepted with probability 
$\min(1,P_\mathrm{eq}(X')/P_\mathrm{eq}(X))$.%
\footnote{ 
 If $\beta_a$ is $\beta_0$ or $\beta_A$, 
 and if the proposed value is not in $\mathcal{A}$, 
 $\beta_a$ is not updated. 
 This procedure ensures the detailed balance.
} 
\end{itemize}
We denote by $\hat{P}_{(1)}$ 
the one-step transition matrix in the $x$ direction, 
which will be repeated $2k$ times, 
and by $\hat{P}_{(2)}$ that in the $\beta$ direction. 
We regard
\begin{align}
 \hat{P}\equiv \hat{P}_{(1)}^k\hat{P}_{(2)}\hat{P}_{(1)}^k
\label{P}
\end{align}
as the transition matrix at a single Markov step. 
One can easily check that $\hat{P}$ satisfies 
the detailed balance condition, 
$\langle X | \hat{P} | X' \rangle \,P_{\rm eq}(X')
=\langle X' | \hat{P} |X \rangle \,P_{\rm eq}(X)$.

With the transition matrix $\hat P$, 
the distance $d_n(X,Y)$ is defined 
for the extended configuration space $\mathcal{M}\times\mathcal{A}$ 
as in \eqref{distance_d_P}:
\begin{align}
  d_n(X,Y) = \sqrt{-\log\biggl(
  \frac{ P_n(X|Y)\, P_n(Y|X)}{P_n(X|X)\, P_n(Y|Y)} \biggr)},
\end{align}
where $P_n(X|Y) = \langle X | \hat{P}^n | Y \rangle$. 
In appendix~\ref{sec:triangle_ineq_ext_coarse}, 
we demonstrate that the distance $d_n(X,Y)$ also satisfies 
the triangle inequality 
on the extended, {\em coarse-grained} configuration space 
$\bar{\mathcal{M}}\times\mathcal{A}$.

\subsection{Matrix elements of the transition matrix}
\label{sec:u_approx}

In this subsection, 
we write down the matrix elements of 
$\hat P=\hat P_{(1)}^k\hat P_{(2)}\hat P_{(1)}^k$ 
for $\beta_a,\beta_b\gg 1$:
\begin{align}
 \langle \mathbf{x},\beta_a\,|\,\hat{P}\,|\,\mathbf{y},\beta_b \rangle
 &= \langle \mathbf{x},\beta_a\,|\, \hat P_{(1)}^k\hat P_{(2)}\hat P_{(1)}^k 
 \,|\,\mathbf{y},\beta_b \rangle
\nonumber
\\
 &= \int_\mathcal{M}  d^D x_1 d^D x_2\, \sum_{a_1,a_2}
 \langle \mathbf{x},\beta_a\,|\,\hat{P}_{(1)}^k 
 \,|\,\mathbf{x}_1,\beta_{a_1}\rangle \times
\nonumber
\\
 &~~~~\times\langle \mathbf{x}_1,\beta_{a_1}\,|\,
 \hat{P}_{(2)}\,|\,\mathbf{x}_2,\beta_{a_2}\rangle \,
 \langle \mathbf{x}_2,\beta_{a_2}\,|\,\hat{P}_{(1)}^k\,|\,
 \mathbf{y},\beta_b\rangle.
\label{P_sim_temp}
\end{align}
We first recall that $k$ is taken to be sufficiently large 
such that the transition in the $x$ direction, 
$\hat P_{(1)}^k$, makes the system be well in local equilibrium. 
Thus, 
configurations will get distributed around local minima 
after the action of $\hat P_{(1)}^k$, 
\begin{align}
 \langle \mathbf{x},\beta_a\,|\,
 \hat{P}_{(1)}^k\,|\,\mathbf{y},\beta_{b}\rangle
 \simeq  P_\mathrm{eq}^{(\mathrm{loc})}(\mathbf{x},\beta_a)\,
 \delta_{[\mathbf{x}]\,[\mathbf{y}]}\,\delta_{a b}
 \quad (\beta_a,\beta_b\gg 1).
\label{P1k}
\end{align}
Here, $[\mathbf{x}]$ represents the local minimum (a lattice point) 
within the mode to which $\mathbf{x}$ belongs. 
The probability distribution of local equilibrium, 
$P_\mathrm{eq}^{(\mathrm{loc})}(\mathbf{x},\beta_a)$,  
should be well approximated by the Gaussian distribution 
around $[\mathbf{x}]$
for large $\beta_a$: 
\begin{align}
  P_\mathrm{eq}^{(\mathrm{loc})}(\mathbf{x},\beta_a)
  \simeq (2\pi\beta_a)^{D/2}\,
  e^{-2\pi^2 \beta_a\,|\mathbf{x}-[\mathbf{x}]|^2}.
\label{Peq}
\end{align}

On the other hand, 
the matrix elements of $\hat{P}_{(2)}$ for $a\neq b$ are given as follows 
(see subsection~\ref{sec:sim_temp_alg}): 
\begin{align}
  &\langle \mathbf{x},\beta_{a}\,|\,\hat{P}_{(2)}
  \,|\,\mathbf{y},\beta_{b}\rangle 
\nonumber
\\
  &=\min\Big(1,\frac{P_\mathrm{eq}
  (\mathbf{x},\beta_{a})}{P_\mathrm{eq}(\mathbf{y},\beta_{b})}\Big)
  \times \frac{1}{2}\,
  (\delta_{a,b+1}+\delta_{a,b-1})\times
  \delta(\mathbf{x}-\mathbf{y}) 
  \quad (a\neq b).
\label{P2}
\end{align}
Substituting \eqref{P1k}--\eqref{P2} to \eqref{P_sim_temp} 
and making some calculation (see appendix \ref{sec:app}), 
we obtain the $a\neq b$ matrix elements 
for $\beta_a,\beta_b\gg 1$: 
\begin{align}
  \label{P_sim_temp2}
  &\langle \mathbf{x},\beta_a\,|\,\hat{P}\,|\,\mathbf{y},\beta_b\rangle\nonumber\\
  &\simeq 
  P_\mathrm{eq}^{(\mathrm{loc})}(\mathbf{x},\beta_a) \times \frac{1}{2} \,
  (\delta_{a,b+1}+\delta_{a,b-1}) \times 
  \int_\mathcal{M} d^D{y}\, 
  \min\Bigl(1,\frac{P_\mathrm{eq}(\mathbf{y},\beta_a)}
  {P_\mathrm{eq}(\mathbf{y},\beta_b)}\Bigr)\,
  P_\mathrm{eq}^{(\mathrm{loc})}(\mathbf{y},\beta_b)\,
  \delta_{[\mathbf{x}]\,[\mathbf{y}]}
\nonumber
\\
 &\simeq 
   P_\mathrm{eq}^{(\mathrm{loc})}(\mathbf{x},\beta_a) \times \frac{1}{2} \,
  (\delta_{a,b+1}+\delta_{a,b-1}) \times 
  \bigl[1- \Delta(\beta_b/\beta_a) \bigr]\,
  \delta_{[\mathbf{x}]\,[\mathbf{y}]} \quad (a \neq b),
\end{align}
where the explicit form of the function $\Delta(z)$ is given 
in \eqref{Delta}. 
The remaining $a=b$ elements can be determined 
from probability conservation, 
and we finally obtain the full matrix elements for $\beta_a,\,\beta_b\gg1$: 
\begin{align}
 \langle \mathbf{x},\beta_a\,|\,\hat{P}\,|\,\mathbf{y},\beta_b\rangle
 &\simeq P_\mathrm{eq}^{(\mathrm{loc})}(\mathbf{x},\beta_a)
 \times
 \biggl[ \frac{1}{2}\,\delta_{a,b+1} [1-\Delta(\beta_{a+1}/\beta_a)]
 +\frac{1}{2}\,\delta_{a,b-1}\, [1-\Delta(\beta_{a-1}/\beta_a)]
\nonumber\\
 &~~~+\frac{1}{2}\,\delta_{a,b}\, 
 [\Delta(\beta_{a+1}/\beta_a)+\Delta(\beta_{a-1}/\beta_a)] \biggr]\,
  \delta_{[\mathbf{x}]\,[\mathbf{y}]}.
\label{P_sim_temp3}
\end{align}

Note that 
the matrix element 
$\langle \mathbf{x},\beta_a\,|\,\hat{P}\,|\,\mathbf{y},\beta_b\rangle$
is factorized to the following form:
\begin{align}
  \langle \mathbf{x},\beta_a\,|\,\hat{P}\,|\,\mathbf{y},\beta_b\rangle
  \simeq P_\mathrm{eq}^{(\mathrm{loc})}(\mathbf{x},\beta_a)\,
  \delta_{[\mathbf{x}]\,[\mathbf{y}]}
  \times \mbox{(function of $R_a$'s)}
  \quad (\beta_a,\beta_b\gg 1),
\end{align}
where $R_a \equiv \beta_a/\beta_{a+1}$ is the ratio of the parameters. 
One can further show that 
$\langle \mathbf{x},\beta_a\,|\,\hat{P}^n\,|\,\mathbf{y},\beta_b\rangle$ 
also takes a factorized form: 
\begin{align}
 \langle \mathbf{x},\beta_a\,|\,\hat{P}^n |\,\mathbf{y},\beta_b\rangle
 \simeq P_\mathrm{eq}^{(\mathrm{loc})}(\mathbf{x},\beta_a)\,
 \delta_{[\mathbf{x}]\,[\mathbf{y}]}
 \times \mbox{(function of $R_a$'s)}
 \quad (\beta_a,\beta_b\gg 1). 
\label{Pn_factorized}
\end{align}
In fact, in the equation 
$\hat{P}^n=\hat{P}_{(1)}^k\hat{P}_{(2)}\hat{P}_{(1)}^k\,\hat{P}^{n-1}$, 
the matrix $\hat{P}_{(1)}^k$ projects $\hat{P}^{n-1}$ 
to the local equilibrium distribution, 
and thus the rest calculation is the same as the case of $n=1$.

\subsection{Geometry of the extended, coarse-grained configuration space}
\label{sec:metric}

We are now in a position to show that 
the extended, coarse-grained configuration space 
$\bar{\mathcal{M}}\times\mathcal{A}$ 
has an asymptotic AdS geometry. 
Here, the metric on $\bar{\mathcal{M}}\times\mathcal{A}$ 
is defined in a similar way to \eqref{distance}:
\begin{align}
 ds^2 = d_n^2(X,\,X+dX),
\end{align}
where $X=(x,\beta_a)$ and $X+dX=(x+dx,\beta_{a+1})$ 
are nearby points in $\bar{\mathcal{M}}\times\mathcal{A}$. 
For the translationally invariant, parity even action \eqref{cosine_action}, 
the metric must take the form 
\begin{align}
  ds^2 = f(\beta)\, d\beta^2 + g(\beta)\, \sum_{\mu=1}^D dx_\mu^2.
\label{metric_sim_temp}
\end{align}

In order to show that 
the metric takes a Euclidean AdS form for large $\beta$, 
we first note that $g(\beta)$ should be an increasing function of $\beta$ 
at least when $\beta$ is large \cite{FMU}. 
In fact, the squared distance $g(\beta)\, \sum_{\mu=1}^D dx_\mu^2$ 
corresponds to those between two different modes for fixed $\beta$, 
and the transition between them should become more difficult 
as $\beta$ increases. 
We then may assume a power-like increase,%
\footnote{ 
 The assumption of power-like increase will be verified numerically 
 in subsection~\ref{sec:numerical}. 
} 
$g(\beta)\propto \beta^q$. 
The exponent $q$ needs to be in the range $0<q<1$ \cite{FMU}. 
In fact, 
$q=1$ when the simulated tempering method is not implemented 
[see \eqref{coarse-grained_metric}], 
and $q$ must be less than this value 
when the simulated tempering is implemented, 
because the distance should be reduced 
by the introduction of the simulated tempering 
and the reduction should be more significant 
for larger $\beta$.

As for $f(\beta)$, 
one can use the expression \eqref{Pn_factorized} 
to calculate the distance along the $\beta$ direction 
for large $\beta$ as follows 
(recall that $R_a=\beta_a/\beta_{a+1}$): 
\begin{align}
  &f(\beta)\,d\beta^2
  =d_n^2((\mathbf{x},\beta_a),(\mathbf{x},\beta_{a+1}))\nonumber\\
  &=-\log\biggl(\frac{
  \langle \mathbf{x},\beta_{a}|\hat{P}^n|\mathbf{x},\beta_{a+1}\rangle\,
  \langle \mathbf{x},\beta_{a+1}|\hat{P}^n|\mathbf{x},\beta_{a}\rangle}
  {\langle \mathbf{x},\beta_{a}|\hat{P}^n|\mathbf{x},\beta_{a}\rangle\,
  \langle \mathbf{x},\beta_{a+1}|\hat{P}^n|\mathbf{x},\beta_{a+1}\rangle}
  \biggr)
\nonumber
\\
  &\simeq
    -\log
    \biggl(
    \frac{
    \bigl[
    P_\mathrm{eq}^{(\mathrm{loc})}(\mathbf{x},\beta_{a})
    \times(\mbox{function of $R_a$'s})
    \bigr]
    \cdot
    \bigl[
    P_\mathrm{eq}^{(\mathrm{loc})}(\mathbf{x},\beta_{a+1})
    \times
    (\mbox{function of $R_a$'s})
    \bigr]
    }
    {\bigl[P_\mathrm{eq}^{(\mathrm{loc})}(\mathbf{x},\beta_{a})\times
    (\mbox{function of $R_a$'s})\bigr]
    \cdot\big[P_\mathrm{eq}^{(\mathrm{loc})}(\mathbf{x},\beta_{a+1})\times
    (\mbox{function of $R_a$'s})\bigr]}\biggr)
\nonumber
\\
  &=-\log\, (\mbox{function of $R_a$'s}).
\label{metric_beta}
\end{align}
Thus, the dependences on local equilibrium distribution disappear 
from the expression, 
only leaving a function that depends on the ratios $R_a$. 
Therefore, $f(\beta)\,d\beta^2$ is invariant under the scaling transformation 
$\beta_a \rightarrow \lambda\,\beta_a$ for large $\beta$, 
and thus we conclude that   
$f(\beta) \propto 1/\beta^2$  
($\beta\gg 1$).

Putting everything together, 
we find that 
the metric on the extended, coarse-grained configuration space 
$\bar{\mathcal{M}}\times\mathcal{A}$ 
takes the following form for $\beta\gg1$:
\begin{align}
  ds^2 = \ell^2\, \Bigl( \frac{d\beta^2}{\beta^2}
  + \alpha\, \beta^q \sum_{\mu=1}^D dx_\mu^2 \Bigr)
  \quad (\ell,\alpha : \mathrm{const} ).
\label{metric_AdS}
\end{align}
Note that this is a Euclidean AdS metric, 
as can be easily seen by a coordinate transformation 
$x_\mu \to (2/q\sqrt{\alpha})\,x_\mu$, 
$\beta \to z^{-2/q}$:
\begin{align}
  ds^2 = \Bigl(\frac{2\ell}{q}\Bigr)^2\cdot\frac{1}{z^2}\,
  \Bigl(dz^2 + \sum_{\mu=1}^D dx_\mu^2\Bigr).
\label{metric_AdS2}
\end{align}




\subsection{Numerical verification of the metric}
\label{sec:numerical}

In this subsection, 
we verify that the geometry of $\bar{\mathcal{M}}\times\mathcal{A}$ 
for large $\beta$ 
is actually given by the metric \eqref{metric_AdS},  
by numerically evaluating the distance $d_n(\mathbf{X},\mathbf{Y})$ 
and comparing the result with a geodesic distance of AdS space
which can be analytically calculated for the metric \eqref{metric_AdS} as  
\begin{align}
  \mathcal{I}(\mathbf{x},\beta_a;\ell,\alpha,q)
  \equiv \frac{4\ell}{q}\, \ln\bigg(
  \frac{ \sqrt{(q\sqrt{\alpha}\,|\mathbf{x}|/4)^2+\beta_a^{-q}}
  + q\sqrt{\alpha}\,|\mathbf{x}|/4}{\beta_a^{-q/2}}
  \bigg). 
\label{AdS_geodesic_distance}
\end{align}

\begin{figure}[t]
  \centering
  \includegraphics[width=5cm]{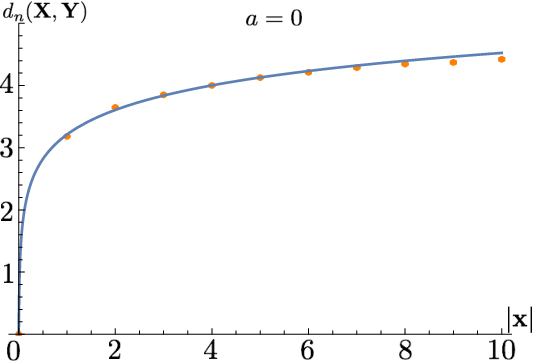}
  \includegraphics[width=5cm]{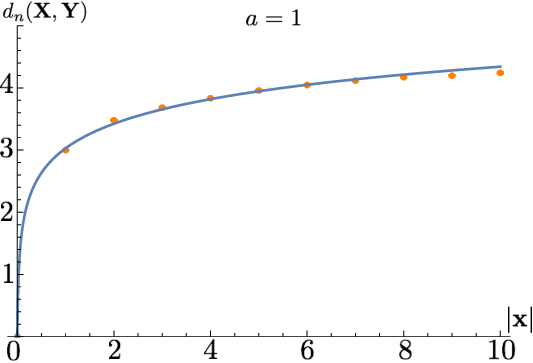}
  \includegraphics[width=5cm]{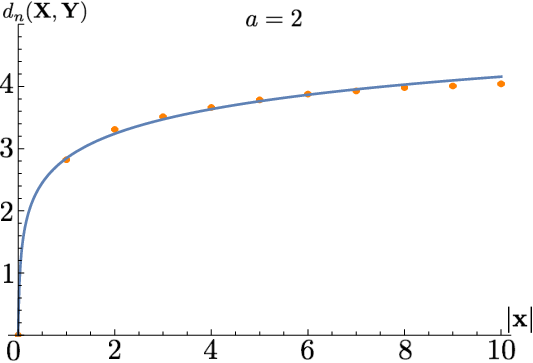}
  \caption{
    $d_n(\mathbf{X}_a,\mathbf{Y}_a)$ 
    for $\beta_a = \exp(10.6 - 2.06a)$.
    $a=0,1,2$ from left to right. 
    The solid line is the geodesic distance \eqref{AdS_geodesic_distance}
    with the fitted parameters.}
  \label{fig:adsfit_opt}
\end{figure}

\begin{figure}[t]
  \centering
  \includegraphics[width=5cm]{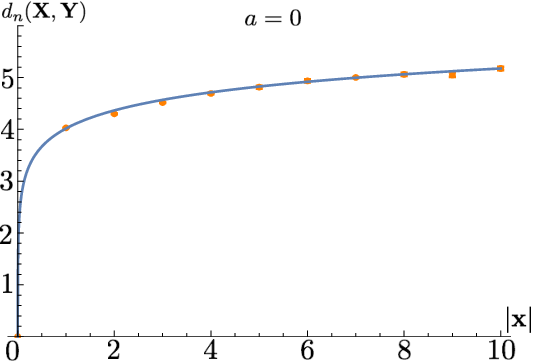}
  \includegraphics[width=5cm]{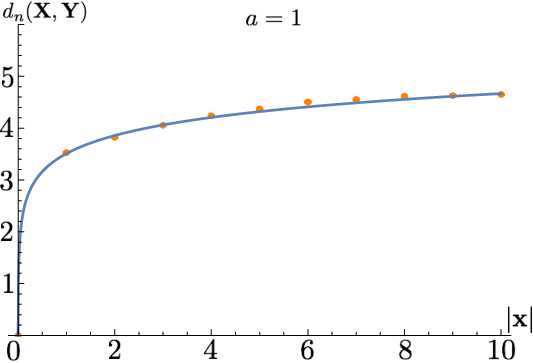}
  \includegraphics[width=5cm]{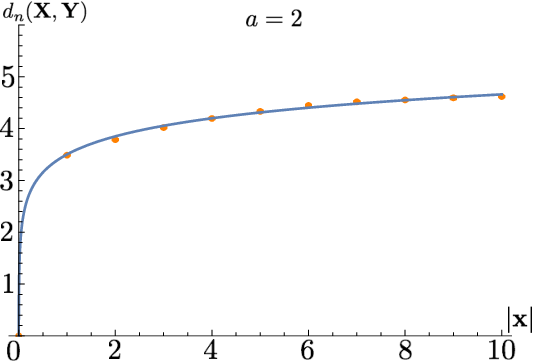}
  \caption{
    $d_n(\mathbf{X}_a,\mathbf{Y}_a)$ 
    for $\beta_a=\exp[ 10.6 - 2.06 a + (-1) ^ a ]$.
    $a=0,1,2$ from left to right. 
    The solid line is the geodesic distance \eqref{AdS_geodesic_distance}
    with the fitted parameters.}
  \label{fig:adsfit_zig}
\end{figure}

We consider a 2-dimensional ($D=2$) configuration space 
that is compactified with the period $2L=100$. 
We set the parameters to $n=100$, $k=8$. 
Transitions in the $\mathbf{x}$ direction at fixed $\beta_a$ 
are generated by the Metropolis algorithm 
using the Gaussian proposal distribution  
with standard deviation 
$\sigma_x(\beta_a)=1.6/(2\pi\sqrt{2\beta_a})$.%
\footnote{
 The values of $\sigma_x$ and $k$ for large $\beta_a$ are set 
 such that $\hat{P}_{(1)}^{2k}$ sufficiently realizes local equilibrium. 
 Note that, when $\beta_a\gg1$, 
 local equilibrium distribution around a local minimum 
 $\mathbf{x}=[\mathbf{x}]$
 can be well approximated by the Gaussian distribution [see \eqref{Peq}]: 
 $
  P_\mathrm{eq}^{(\mathrm{loc})}(\mathbf{x},\beta_a)
  \simeq (2\pi\beta_a)^{D/2}\,
  e^{-2\pi^2 \beta_a\,|\mathbf{x}-[\mathbf{x}]|^2}.
 $
 Therefore, when we propose from a configuration $\mathbf{x}$ 
 a new configuration $\mathbf{x}'=\mathbf{x}+\mathbf{y}$ 
 by using the Gaussian distribution with standard deviation $\sigma_x$, 
 the difference of the action, $\Delta S=S(\mathbf{x}')-S(\mathbf{x})$, 
 appearing in the Metropolis acceptance probability $[\min(1,e^{-\Delta S})]$ 
 has the expectation value 
 $ \langle \Delta S \rangle 
 \simeq 2\pi^2 \beta_a\, \langle \mathbf{y}^2 \rangle 
 \simeq 2\pi^2 \beta_a\, D \sigma_x^2$. 
 We will set this to be an order-one constant $(\equiv \alpha^2/2)$ 
 in order that the acceptance rate is significant 
 and does not change largely for different $\beta_a$'s. 
 This makes $\sigma_x$ depend on $\beta_a$  
 as $\sigma_x(\beta_a)=\alpha/(2\pi\sqrt{D \beta_a})$. 
 In this paper, we set $\alpha=1.6$ and $k=8$, 
 which we confirm to give significant acceptance rates 
 in actual Monte Carlo calculations. 
\label{fn:sigmax}
} 
As for transitions in the $\beta$ direction, 
in order to reduce boundary effects for a finite interval $\{\beta_a\}$, 
we extend the set $\{\beta_a\}$ such that $a$ runs from $-200$ to $200$, 
and only consider the distances between the points 
$\mathbf{X}_a \equiv (\mathbf{x},\beta_a)$ 
and $\mathbf{Y}_a \equiv (\mathbf{0},\beta_a)$ 
with $a=0,1,2$.  
The explicit functional form of $\beta_a=\beta(a)$ are chosen in two ways 
in order to show that the asymptotic AdS geometry is always obtained 
independently of the choice of $\beta_a$'s. 
The first choice is an exponential form $\beta_a = \exp(10.6 - 2.06a)$, 
and the other is a zigzag form $\beta_a=\exp[ 10.6 - 2.06 a + (-1) ^ a ]$. 
The exponential form given above 
is actually the optimized form that minimize distances, 
as will be found in subsection~\ref{sec:num_optim}.  
We generate $5 \times 10^6$ configurations 
for each initial configuration $\mathbf{Y}_a$ $(a=0,1,2)$, 
and then calculate the distance between $\mathbf{X}_a=(\mathbf{x},\beta_a)$ 
and $\mathbf{Y}_a=(\mathbf{0},\beta_a)$ 
for various $\mathbf{x}$ $(|\mathbf{x}|=0,\ldots,10)$.  
The obtained results for the first choice are depicted 
as dots in figure~\ref{fig:adsfit_opt}, 
while those for the second choice in figure~\ref{fig:adsfit_zig}.
%
%
%

Using the obtained numerical results, 
we fit the parameters $\ell,\,\alpha,\,q$ in \eqref{metric_AdS} 
by minimizing the following $\chi^2$:
\begin{align}
 \chi^2(\ell,\alpha,q)
 \equiv \sum_{\mathbf{x},a}
 \frac{\bigl[d_n\bigl((\mathbf{x},\beta_a),(\mathbf{0},\beta_a)\bigr)
 - \mathcal{I}(\mathbf{x},\beta_a;\ell,\alpha,q)\bigr]^2}
 {\mbox{Var}\bigl[d_n\bigl((\mathbf{x},\beta_a),(\mathbf{0},\beta_a)\bigr)\bigr]}.
\end{align} 
For the first choice (exponential form) 
we obtain $\ell = 0.0445 \pm 0.0002$, 
$\alpha = (1.180 \pm 0.022 ) \times 10^5$, 
$q = 0.313 \pm 0.001$, with $\sqrt{\chi^2/30} = 13.6$, 
and for the second choice (zigzag form) 
we obtain $\ell = 0.0624 \pm 0.0002$, 
$\alpha = (4.421 \pm 0.165) \times 10^5$, 
$q = 0.497 \pm 0.002$, with $\sqrt{\chi^2/30} = 6.9$. 
%
We draw the geodesic distances with these fitted parameters 
as solid lines in figs.~\ref{fig:adsfit_opt} and \ref{fig:adsfit_zig}. 
The good agreement shows that the distances can be regarded 
as geodesic distances of an asymptotically Euclidean AdS metric, 
and also that this conclusion does not depend 
on the choice of the functional form of $\beta_a$.

\section{Optimization of the tempering parameter}
\label{sec:optim}

In this section, 
we optimize the tempering parameter 
by demanding that the distances get minimized. 
We will show that a simple, geometrical consideration gives the conclusion 
that $\beta_a$'s take an exponential form for large $\beta_a$. 
We further will confirm this conclusion 
by numerically optimizing the tempering parameter set. 
We show that the optimized parameter set actually takes an exponential form 
for large $\beta_a$ 
and that they also exhibit a horizon for small $\beta_a$, 
beyond which configurations can move freely in the $x$ direction.

\subsection{Need for the optimization of $\beta_a$}
\label{sec:need_optim}

The tempering parameter set $\mathcal{A}=\{\beta_a\}$ must be chosen 
such that transitions in the $\beta$ direction have significant acceptance rates 
and also that transitions in the $x$ direction are easy 
for some $\beta_a$. 
If $\beta$ represents the overall coefficient of the action 
and we order $\mathcal{A}=\{\beta_a\}$ 
as $\beta_0 > \beta_1 > \cdots > \beta_A$ 
[as we have done for the action \eqref{cosine_action}], 
then the above requirement means 
that $\beta_a$ must be placed densely to some extent 
(and thus $A$ must be large), 
and also that $\beta_a$ $(a\sim A)$ must be sufficiently small 
such that severe potential barriers no longer exist there. 
However, if one introduced too many $\beta_a$'s, 
the size of the subsample with $\beta_{a=0}$ would become small 
and the sample average would get a large statistical error. 
This enforces us to make a careful adjustment 
of the elements of the set $\mathcal{A}=\{\beta_a\}$. 
We will show in the next subsection   
that this adjustment can be carried out in a simple, geometrical way.

\subsection{Geometrical optimization of the tempering parameter}
\label{sec:geom_optim}

In section~\ref{sec:emergence2}, 
we showed that the metric in the extended, coarse-grained configuration space 
has the component: 
\begin{align}
 ds^2|_{x={\rm const}} = d_n^2\bigl((x,\beta),(x,\beta+d\beta)\bigr)
 = {\rm const.}\,\frac{d\beta^2}{\beta^2}
 = {\rm const.}\,\Bigl(\frac{\beta'(a)}{\beta(a)}\Bigr)^2\,da^2,
\end{align}
where we regard $\beta(a) \equiv \beta_a$ as a continuous function of $a$. 
To accelerate the relaxation of the probability distribution, 
we need to optimize the tempering parameter set $\{\beta_a\}$. 
Since almost all of the transitions occur only in the $\beta$ direction 
for large $\beta_a$, 
the optimization should be made 
so that the transitions in the $\beta$ direction has no obstacle 
for large $\beta$. 
Since it is the parameter $a$ that is directly dealt with by a MCMC simulation, 
the above requirement can be re-expressed as a geometrical statement 
that the metric in the $a$ direction is flat. 
This means the equality
\begin{align}
 \frac{\beta'(a)}{\beta(a)} = {\rm const.},
\end{align}
and thus we conclude that the optimized tempering parameter set 
should have an exponential form, 
$\beta_a = \beta_0\,R^{-a}$ ($R$: constant). 
Note that this form is expected only for large $\beta_a$'s, 
because for small enough $\beta_a$, 
transitions in the $x$ direction come to occur frequently 
and thus the above argument no longer holds.

\subsection{Numerical confirmation of the geometrical optimization}
\label{sec:num_optim}

In this subsection, 
we numerically optimize the tempering parameter set $\{\beta_a\}$, 
and show that the optimized set actually takes an exponential form 
for large $\beta_a$. 
We also show that there exists a horizon for small $\beta_a$, 
beyond which configurations can move freely in the $x$ direction.

Below is the numerical algorithm we take 
to optimize the tempering parameter set. 
We here fix 
$\beta_0$ (the overall coefficient of the original action), 
$n$ (the number of steps), 
and $A$ (the number of the additional parameters), 
and vary the parameters 
$\bm{\beta} \equiv \{\beta_a\}_{a=1,\ldots,A}$ 
such that the distances between different modes 
$\mathbf{X}_1=(\mathbf{x}_1,\beta_0),\,\mathbf{X}_2=(\mathbf{x}_2,\beta_0)
 \in \bar{\mathcal{M}}\times\mathcal{A}$ 
are minimized. 
For given initial parameters $\bm{\beta}$, 
we update them with the following steps:%
\footnote{
 To ensure transitions in the $\beta$ direction to occur sufficiently, 
 $n$ should be set to be larger than $O(A^2)$, 
 as can be estimated by regarding the transitions as pure random walks. 
 At the same time, however,
 $n$ should not be taken too large 
 in order to avoid a large accumulation of boundary effects.
 In the following calculation we will set $n$ to $n\sim 1.5 A^2$. 
} 
\begin{enumerate}

\item
 We calculate the distance $d_n(\mathbf{X}_1,\mathbf{X}_2;\bm{\beta})$ 
 for $\bm{\beta}$.%
\footnote{
 We here denote the distance by $d_n(\mathbf{X}_1,\mathbf{X}_2;\bm{\beta})$ 
 to specify its dependence on the tempering parameter set 
 $\bm{\beta}=\{\beta_a\}$.
} 

\label{item:1}

\item
 We propose a new parameter set 
 $\bm{\beta}'= \{ \beta_0,\ldots,\beta_{a-1},\beta_a',\beta_{a+1},
 \ldots,\beta_A\}$
 by generating $\ln{\beta_a'}$ using the Gaussian distribution 
 with mean $\ln\beta_{a}$ and standard deviation $\sigma_\beta$ 
 (a constant given in advance) for randomly selected $a$. 
 If the obtained parameters are not ordered, 
 we repeatedly generate another set until we get ordered ones. 
\label{item:2}
 
\item 
 We calculate the distance $d_n(\mathbf{X}_1,\mathbf{X}_2;\bm{\beta}')$
 for the new parameter set, 
 and update $\bm{\beta}$ to $\bm{\beta'}$ 
 if $d_n(\mathbf{X}_1,\mathbf{X}_2;\bm{\beta}')
 < d_n(\mathbf{X}_1,\mathbf{X}_2;\bm{\beta})$. 
\label{item:3}

\item
 We repeat steps \ref{item:2} and \ref{item:3}. 
 Since the calculated distance can be unintentionally small 
 due to a statistical error, 
 it can happen that we reject a set which should be accepted. 
 To avoid this to happen frequently, 
 we recalculate the distance with a larger precision 
 when proposed sets are rejected $N_\mathrm{discard}$ times sequentially 
 ($N_\mathrm{discard}$ is a number given in advance).
\label{item:4}
\end{enumerate}

\begin{figure}[t]
  \centering
  \includegraphics[width=9cm]{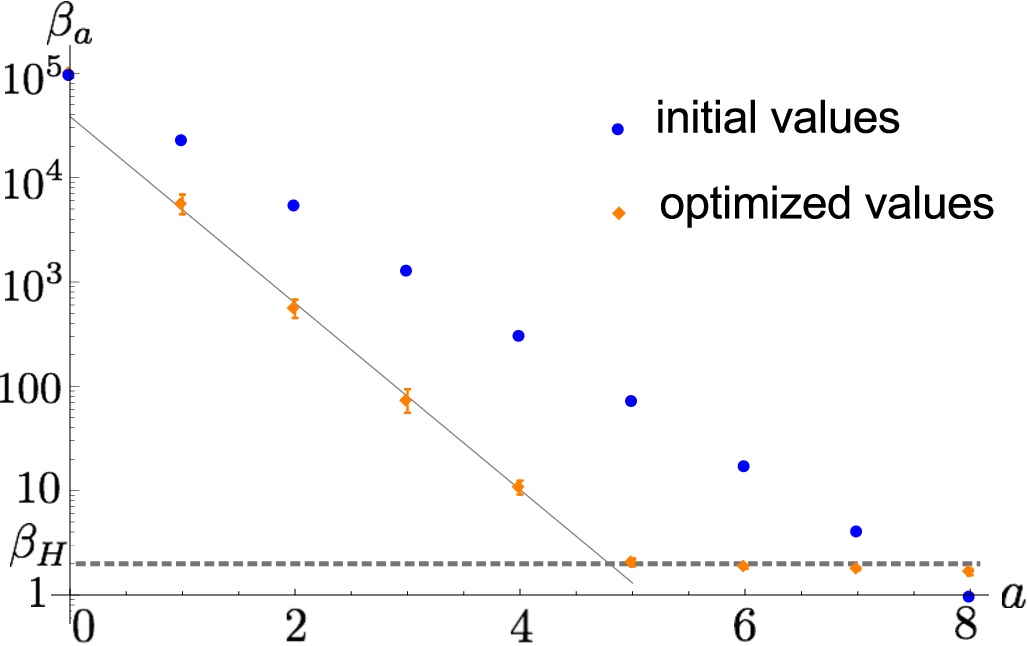}
  \caption{
  Optimized values for $\bm{\beta}=\{\beta_a\}$ $(a=1,\ldots,8)$.
  The blue dots are the initial values, 
  and the orange dots give the resulting optimized values 
  that minimize the distance $d_n(\mathbf{X}_1,\mathbf{X}_2)$. 
  The horizontal dashed line at $\beta_H\simeq 2$ 
  represents the horizon, 
  beyond which configurations can move freely 
  in the $x$ direction.
  }
  \label{fig:beta_both}
\end{figure}

Figure~\ref{fig:beta_both} shows the optimized values of 
$\{\beta_a\}$ $(a=1,\ldots,A)$ 
for a 2-dimensional ($D=2$) noncompact configuration space 
with the parameters $k=8$ and $n=100$. 
%
We set $\beta_0=10^5$, $A=8$, $\sigma_\beta=(\ln\beta_0)/(4A)$ 
and $N_\mathrm{discard}=10$. 
We start from initial values $\beta_a = \beta_0^{1-a/A}$. 
We use the Metropolis algorithm to generate a transition 
in the $\mathbf{x}$ direction, 
using the Gaussian distribution with standard deviation 
$\sigma_x(\beta_a)=1.6/(2\pi\sqrt{2\beta_a})$ as a proposal distribution. 
We calculate the distance $d_{n=100}(\mathbf{X}_1,\mathbf{X}_2;\bm{\beta})$ 
between $\mathbf{X}_1=((1,0),\beta_0)$ 
and $\mathbf{X}_2=((0,0),\beta_0)$ 
for given $\bm{\beta}$. 
We first repeat steps \ref{item:2} and \ref{item:3} 10,000 times, 
where at least 100 paths from $\mathbf{X}_1$ to $\mathbf{X}_2$ 
are generated to calculate the distance. 
Following this, 
we then repeat the same procedures 15,000 times, 
but this time at least 200 paths are generated. 
After we obtain the optimized values 
(drawn as orange points in fig.~\ref{fig:beta_both}), 
we fit the points in the region $1 \leq a \leq 4$ 
with a linear function $c_1 a + c_2$ by minimizing $\chi^2$. 
We obtain $c_1 = -2.06 \pm 0.08$, $c_2 = 10.6 \pm 0.2$ 
with $\sqrt{\chi^2/4}=0.48$. 
The points near the boundary $a=0$ behaves differently 
which may be explained as boundary effects. 
We also see that $\beta_a$ $(a=5,6,7,8)$ have almost the same value, 
$\beta_H\sim 2$. 
$\beta_H$ represents a ``horizon'', 
beyond which the potential barriers in the $x$ direction disappear 
and configurations can move freely in the direction.%
\footnote{ 
 The position of the horizon, $\beta_H$, depends on the algorithm, 
 especially on the standard deviation $\sigma_x$ 
 of the Gaussian proposal for transitions in the $x$ direction. 
 This horizon differs from the horizon of a Euclidean blackhole 
 where a single $S^1$-cycle vanishes. 
} 
Thus, the value $A=8$ is actually too large, 
and we find that five parameters $\beta_a$ $(a=1,\ldots,5)$ 
should be enough for this calculation.

We thus arrive at the conclusion that 
the optimized tempering parameter is given by an exponential form, 
$\beta_a = \beta_0\,R^{-a}$,  
except for those near boundaries. 
This confirms the geometrical optimization 
made in subsection~\ref{sec:geom_optim}.

\begin{figure}[t]
  \centering
  \includegraphics[width=8cm]{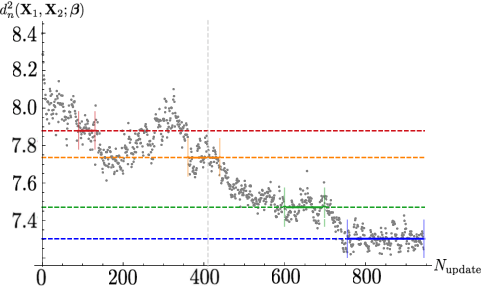}~
  \includegraphics[width=8cm]{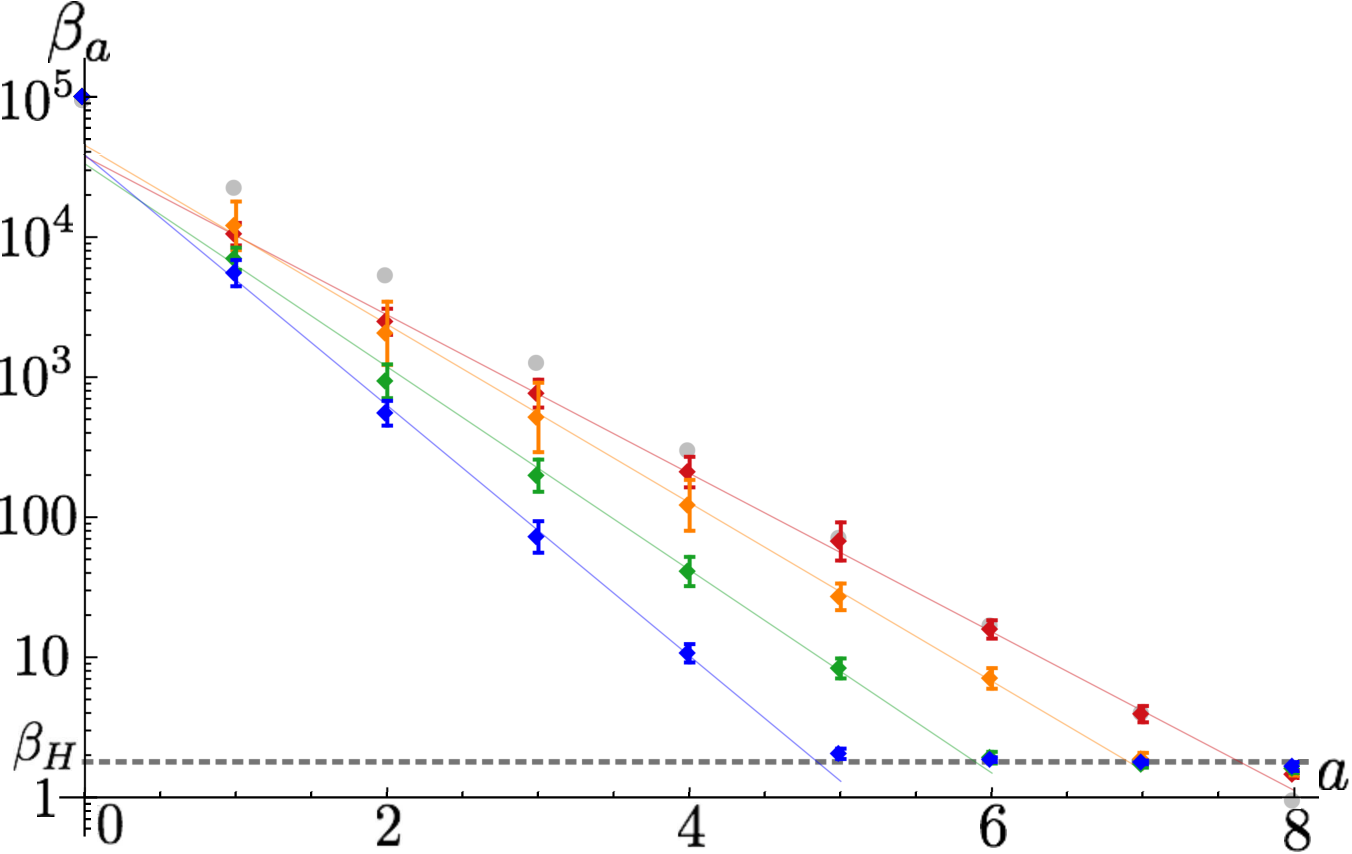}
  \caption{
  (Left) History of $d_n^2(\mathbf{X}_1,\mathbf{X}_2;\bm{\beta})$ 
  with respect to the updates of $\bm{\beta}$. 
  Smaller $\sigma_\beta$ is used for step 3 
  after the number of updates of $\bm{\beta}$ exceeds 410.
  (Right) Metastable solutions. 
  Each solid line represents a metastable solution $\bm{\beta}=\{\beta_a\}$
  defined as the average of the values of $\bm{\beta}$'s 
  belonging to a segment drawn in the left panel.
  }
  \label{fig:metastable_d2}
\end{figure}

We end this subsection with a comment that 
there are actually many metastable solutions for $\bm{\beta}$ 
and all such solutions also take exponential forms. 
The left panel of fig.~\ref{fig:metastable_d2} is 
the history of $d_n^2(\mathbf{X}_1,\mathbf{X}_2;\bm{\beta})$ 
with respect to the updates of $\bm{\beta}$. 
%
We there see several plateaus that correspond to metastable states. 
The right panel shows that the metastable solutions, 
each corresponding to a plateau, 
actually take exponential forms.

\section{Conclusion and outlook}
\label{sec:conclusion}

In this paper, we have investigated the global geometry of a stochastic system 
whose equilibrium distribution is highly multimodal 
with a large number of degenerate vacua 
that are distributed uniformly.

We implemented the simulated tempering algorithm to such a system 
by taking the tempering parameter 
to be the overall coefficient $\beta_0$ of the action. 
We showed that 
the geometry of the extended, coarse-grained configuration space 
$\bar{\mathcal{M}}\times\mathcal{A}$ 
is given by an asymptotically AdS space. 
With this knowledge of geometry, 
we deduced the conclusion that 
the optimized tempering parameter set $\{\beta_a\}$ 
must take an exponential form $\beta_a = \beta_0\,R^{-a}$ for large $\beta_a$. 
This conclusion was confirmed by numerical calculations, 
where it was also found that a horizon exists at small $\beta$, 
around which transitions in the $x$ direction become easy.

As for the future work, 
it would be interesting to investigate the large scale geometry 
of the configuration space in the Yang-Mills theory, 
especially by coarse-graining the configuration space 
according to the topological charges.

It should also be important to construct a more convenient method 
to find the optimized parameters for a given MCMC algorithm. 
For the simulated tempering algorithm, for example, 
the optimization is to determine the functional form of $\beta_a=\beta(a)$. 
Thus, it should be nice 
if we can find a local functional of $\beta(a)$ 
such that the optimized form is directly given 
by solving its Euler-Lagrange equation, 
and further if this local functional has some relationship 
with the Einstein-Hilbert action. 
We expect that this investigation may also give a clue to
understanding the hidden stochastic character in general relativity.

In this paper, we have discussed only the case 
where the action $S(x)$ is real. 
When the action takes complex values 
as in QCD at finite density 
or in the quantum Monte Carlo computation 
of the Hubbard model away from half filling, 
we cannot directly use the present definition of distance  
because the Boltzmann weight is complex-valued. 
It must be important to investigate 
whether nice distances can also be introduced to these algorithms. 
In particular, it must be interesting 
to investigate the large scale geometry of 
the (generalized) Lefschetz thimble method 
\cite{Cristoforetti:2012su,Cristoforetti:2013wha,Mukherjee:2013aga,
Fujii:2013sra,Cristoforetti:2014gsa,
Alexandru:2015sua,Fukuma:2017fjq,Alexandru:2017oyw} 
with the implementation of the tempering algorithm 
that takes the tempering parameter to be the flow time 
of the antiholomorphic gradient flow 
\cite{Fukuma:2017fjq,Alexandru:2017oyw}.

A study along these lines is now in progress and will be reported elsewhere.

\section*{Acknowledgments}
The authors thank Hikaru Kawai, So Matsuura, 
Jun Nishimura and Asato Tsuchiya 
for useful discussions. 
This work was partially supported by JSPS KAKENHI 
(Grant Numbers 16K05321, 18J22698 and JP17J08709).

\appendix

\section{Triangle inequality on the extended, coarse-grained configuration space}
\label{sec:triangle_ineq_ext_coarse}

We demonstrate that the distance $d_n(X,Y)$ satisfies 
the triangle inequality 
also in the extended, coarse-grained configuration space 
$\bar{\mathcal{M}}\times\mathcal{A}$ for $\beta_a \gg 1$.

For this purpose, we explicitly calculate the value 
$d_n(X,Y)+d_n(Y,Z)-d_n(X,Z)$ by taking three points 
$X,Y,Z \in\bar{\mathcal{M}}\times\mathcal{A}$ in various ways. 
We use the action~\eqref{cosine_action} with parameters $D=2$ and $L=100$. 
The tempering parameter is set to 
$\beta_a=100\cdot10^{-3a/10}$ ($a=-200,\ldots,200$).%
\footnote{ 
  We have extended the region of $\{a\}$ 
  in order to reduce boundary effects. 
  See also discussions in subsection~\ref{sec:numerical}. 
} 
To generate a transition in the $\mathbf{x}$ direction, 
we use the same Metropolis algorithm 
but now set the standard deviation of the Gaussian proposal distribution 
to $\sigma_x(\beta_a)=1.6/(2\pi\sqrt{2\beta_a})$. 
We set $k=8$ and $n=1,000$, 
and generate $10^6$ configurations from each initial configuration. 
In table~\ref{tab:triangle}, 
we show the result for $X=((0,0),\beta_0)$, $Y=((y,0),\beta_a)$ 
and $Z=((3,0),\beta_1)$ with $y=0,\ldots,6$ and $a=-2,\ldots,2$. 
%
\begin{table}[t]
  \centering
  \begin{tabular}{|c|c|c|c|c|c|c|c|}\hline
    $a$ \textbackslash ~$y$&0&1&2&3&4&5&6 \\\hline
    -2&0.0$\pm$1.2&1.4$\pm$0.1&2.3$\pm$0.1&0.1$\pm$1.0&2.7$\pm$0.1&3.1$\pm$0.1&3.5$\pm$0.1 \\\hline
    -1&0.1$\pm$0.2&2.3$\pm$0.1&2.3$\pm$0.1&0.2$\pm$0.1&2.6$\pm$0.1&3.2$\pm$0.1&3.5$\pm$0.1 \\\hline
    0&0.0$\pm$0.1&2.2$\pm$0.1&2.2$\pm$0.1&0.2$\pm$0.2&2.7$\pm$0.1&3.3$\pm$0.2&3.4$\pm$0.1 \\\hline
    1&0.1$\pm$0.2&2.1$\pm$0.1&2.1$\pm$0.1&0.0$\pm$0.1&2.5$\pm$0.1&3.0$\pm$0.1&3.4$\pm$0.2 \\\hline
    2&0.11$\pm$0.16&2.1$\pm$0.1&2.1$\pm$0.1&0.13$\pm$0.15&2.4$\pm$0.1&3.0$\pm$0.1&3.3$\pm0.1$ \\\hline
  \end{tabular}
  \caption{$d_n(X,Y)+d_n(Y,Z)-d_n(X,Z)$ 
  for $X=((0,0),\beta_0)$, $Y=((y,0),\beta_a)$ and $Z=((3,0),\beta_1)$.
  $\beta_a$ are set to $\beta_{-2}\simeq 400$, $\beta_{-1}\simeq 200$, 
  $\beta_{0} = 100$, $\beta_{1}\simeq 50$, $\beta_{2}\simeq 25$. 
  All the entries are positive within statistical errors, 
  which shows that the triangle inequality holds.}
  \label{tab:triangle}
\end{table}
We see that $d_n(X,Y)$ does satisfy the triangle inequality 
on $\bar{\mathcal{M}}\times\mathcal{A}$.

\section{Calculation of eq.~(\ref{P_sim_temp2})}
\label{sec:app}

In this appendix, we evaluate the integral 
that appears in \eqref{P_sim_temp2}: 
\begin{align}
  I \equiv
  \int_\mathcal{M} d^Dx\, \min\Bigl(1,\frac{P_\mathrm{eq}(\mathbf{x},\beta_b)}
  {P_\mathrm{eq}(\mathbf{x},\beta_a)}\Bigr)\,
  P_\mathrm{eq}^{(\mathrm{loc})}(\mathbf{x},\beta_a).
\label{app:integral}
\end{align}
We are concerned with the case where $\beta_a,\,\beta_b\gg 1$, 
and thus can approximate $P_\mathrm{eq}(\mathbf{x},\beta_a)$ 
by the Gaussian distributions around local minima [see \eqref{Peq}]. 
Due to the translational invariance, 
we only need to consider the local minimum at the origin, 
for which the integrand becomes 
\begin{align}
  \min\Big((2\pi\beta_a)^{D/2}\,e^{-2\pi^2 \beta_a r^2}, 
  (2\pi\beta_b)^{D/2}\,e^{-2\pi^2 \beta_b r^2}\Big) 
  \quad (r^2\equiv \mathbf{x}^2).
\label{integrand}
\end{align}
Let us first consider the case $\beta_a < \beta_b$. 
%
\begin{figure}[t]
  \centering
  \includegraphics[width=7cm]{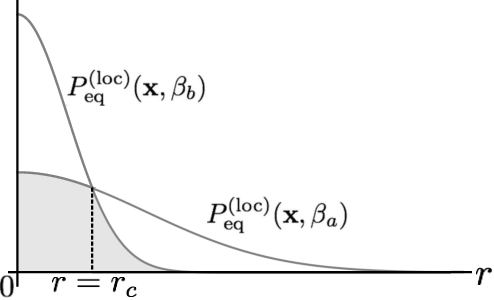}
  \caption{
  The integrand \eqref{integrand} for the case $\beta_a < \beta_{b}$. 
  The integral~(\ref{app:integral}) is given by the volume of the shaded region.}
  \label{fig:rc}
\end{figure}
The integral of (\ref{integrand}) then can be evaluated as 
(see Fig.~\ref{fig:rc})
\begin{align}
  I &\simeq \Omega_{D-1}\,
  \bigg[ \int_0^{r_c} dr \ r^{D-1} (2\pi \beta_a)^{D/2}\,
  e^{-2\pi^2 \beta_a r^2}
  + \int_{r_c}^\infty dr \  r^{D-1} (2\pi \beta_b)^{D/2}\,
  e^{-2\pi^2 \beta_b r^2} \bigg]
\nonumber
\\
  &= 1 - \Omega_{D-1}\,
  \bigg[ -(2\pi \beta_a)^{D/2} \int_0^{r_c} dr \ r^{D-1}\,
  e^{-2\pi^2 \beta_a r^2}
  + (2\pi \beta_b)^{D/2} \int_0^{r_c} dr \  r^{D-1} e^{-2\pi^2 \beta_b r^2}
 \bigg], 
\label{app:integral2}
\end{align}
where $\Omega_{D-1} = 2\pi^{D/2} / \Gamma(D/2) $ is the volume of 
$(D-1)$-sphere, and $r_c$ is defined by
\begin{align}
  r_c \equiv
 \sqrt{ \frac{(D/2) \ln(\beta_b/\beta_a)}{ 2\pi^2 (\beta_b-\beta_a) } }.
\label{app:rc}
\end{align}
The remaining integral in \eqref{app:integral2} can be written 
in terms of the lower incomplete gamma function 
$\gamma(z,p) = \int_0^p dt \, e^{-t}\, t^{z-1}$ 
as 
\begin{align}
 I&\simeq 1
 - \frac{1}{\Gamma(D/2)}\,\big[-\gamma(D/2,{\tilde{r}_c}^2)
 + \gamma(D/2,{\tilde{s}_c}^2)\big],
\label{app:integral3}
\end{align}
where ${\tilde{r}_c}$ and ${\tilde{s}_c}$ are defined by 
\begin{align}
  \tilde{r}_c&\equiv r_c\sqrt{2\pi^2\beta_a} 
  = \sqrt{ 2\pi^2 \frac{(D/2) \ln(\beta_b/\beta_a)}
  { 2\pi^2 (\beta_b/\beta_a- 1) } }, 
\nonumber
\\
  \tilde{s}_c&\equiv r_c\sqrt{2\pi^2\beta_b}
  = \sqrt{ 2\pi^2 \frac{(D/2) \ln(\beta_b/\beta_a)}
  { 2\pi^2 (1-\beta_a/\beta_b) } },
\label{app:rc2}
\end{align}
and depend on $\beta_a,\,\beta_b$ 
only through the ratio $\beta_b/\beta_a$. 
The integral $I$ for the case $\beta_a > \beta_b$ 
can also be evaluated in the same way. 
We thus obtain 
\begin{align}
  I\simeq 1
  - \frac{1}{\Gamma(D/2)}\,\bigl|\gamma(D/2,{\tilde{r}_c}^2)
  - \gamma(D/2,{\tilde{s}_c}^2)\bigr| 
  \equiv 1- \Delta(\beta_b/\beta_a), 
\label{Delta}
\end{align}
which is actually an expression 
valid for the both cases, $\beta_a \lessgtr \beta_b$.

\baselineskip=0.9\normalbaselineskip




\begin{thebibliography}{99}
\setlength{\itemsep}{-2pt}

\bibitem{FMU} 
  M.~Fukuma, N.~Matsumoto and N.~Umeda,
  ``Distance between configurations in Markov chain Monte Carlo simulations,''
  JHEP {\bf 1712}, 001 (2017)
  [arXiv:1705.06097 [hep-lat]].

\bibitem{Creutz:1987xi} 
  M.~Creutz,
  ``Overrelaxation and Monte Carlo Simulation,''
  Phys.\ Rev.\ D {\bf 36}, 515 (1987).
  
\bibitem{Marinari:1992qd} 
  E.~Marinari and G.~Parisi,
  ``Simulated tempering: A New Monte Carlo scheme,''
  Europhys.\ Lett.\  {\bf 19}, 451 (1992)
  [hep-lat/9205018].

\bibitem{Swendsen1986}
  R.~H.~Swendsen and J.-S.~Wang,
  ``Replica Monte Carlo Simulation of Spin-Glasses,''
  Phys.\ Rev.\ Lett.\ {\bf 57} 2607 (1986). 

\bibitem{Geyer1991}
  C.~J.~Geyer, 
  ``Markov Chain Monte Carlo Maximum Likelihood,''
  in Computing Science and Statistics: Proceedings of the 23rd Symposium
  on the Interface, American Statistical Association, New York, p.~156 (1991).

\bibitem{Earl2005}
  D.~J.~Earl and M.~W.~Deem, 
  ``Parallel tempering: Theory, applications, and new perspectives,''
  Phys.\ Chem.\ Chem.\ Phys.\ {\bf 7}, 3910 (2005).

\bibitem{Luscher:2010iy} 
  M.~L\"{u}scher,
  JHEP {\bf 1008}, 071 (2010)
  Erratum: [JHEP {\bf 1403}, 092 (2014)]
  doi:10.1007/JHEP08(2010)071, 10.1007/JHEP03(2014)092
  [arXiv:1006.4518 [hep-lat]].
  
\bibitem{Cristoforetti:2012su} 
  M.~Cristoforetti, F.~Di Renzo and L.~Scorzato,
  ``New approach to the sign problem in quantum field theories: 
  High density QCD on a Lefschetz thimble,''
  Phys.\ Rev.\ D {\bf 86}, 074506 (2012)
  [arXiv:1205.3996 [hep-lat]].

\bibitem{Cristoforetti:2013wha} 
  M.~Cristoforetti, F.~Di Renzo, A.~Mukherjee and L.~Scorzato,
  ``Monte Carlo simulations on the Lefschetz thimble: Taming the sign problem,''
  Phys.\ Rev.\ D {\bf 88}, no. 5, 051501 (2013)
  [arXiv:1303.7204 [hep-lat]].
 
\bibitem{Mukherjee:2013aga} 
  A.~Mukherjee, M.~Cristoforetti and L.~Scorzato,
  ``Metropolis Monte Carlo integration on the Lefschetz thimble: 
  Application to a one-plaquette model,''
  Phys.\ Rev.\ D {\bf 88}, no. 5, 051502 (2013)
  [arXiv:1308.0233 [physics.comp-ph]].
  
\bibitem{Fujii:2013sra} 
  H.~Fujii, D.~Honda, M.~Kato, Y.~Kikukawa, S.~Komatsu and T.~Sano,
  ``Hybrid Monte Carlo on Lefschetz thimbles
  - A study of the residual sign problem,''
  JHEP {\bf 1310}, 147 (2013)
  [arXiv:1309.4371 [hep-lat]].

\bibitem{Cristoforetti:2014gsa} 
  M.~Cristoforetti, F.~Di Renzo, G.~Eruzzi, A.~Mukherjee, 
  C.~Schmidt, L.~Scorzato and C.~Torrero,
  ``An efficient method to compute the residual phase on a Lefschetz thimble,''
  Phys.\ Rev.\ D {\bf 89}, no. 11, 114505 (2014)
  [arXiv:1403.5637 [hep-lat]].

\bibitem{Alexandru:2015sua} 
  A.~Alexandru, G.~Ba\c sar, P.~F.~Bedaque, G.~W.~Ridgway and N.~C.~Warrington,
  ``Sign problem and Monte Carlo calculations beyond Lefschetz thimbles,''
  JHEP {\bf 1605}, 053 (2016)
  [arXiv:1512.08764 [hep-lat]].

\bibitem{Fukuma:2017fjq} 
  M.~Fukuma and N.~Umeda,
  ``Parallel tempering algorithm for integration over Lefschetz thimbles,''
  PTEP {\bf 2017}, no. 7, 073B01 (2017)
  [arXiv:1703.00861 [hep-lat]].

\bibitem{Alexandru:2017oyw} 
  A.~Alexandru, G.~Ba\c sar, P.~F.~Bedaque and N.~C.~Warrington,
  ``Tempered transitions between thimbles,''
  Phys.\ Rev.\ D {\bf 96}, no. 3, 034513 (2017)
  [arXiv:1703.02414 [hep-lat]].

\end{thebibliography}
\end{document}